\documentstyle[12pt]{article}
\begin{document}
\thispagestyle{empty}
\begin{center}
{\LARGE \tt \bf Torsion and Quantized Vortices}
\end{center}

\vspace{0.5cm}

\begin{center}
{\large \tt L.C. Garcia de Andrade\footnote{Grupo de Relatividade e
Teoria de Campo, Instituto de F\'{\i}sica, UERJ} and C. Sivaram\footnote{
Indian Institute of Astrophysics, Bangalore, India and Istituto di 
F\'{\i}sica dell'a Universita', Ferrara, Italia.}}
\end{center}

\vspace{2cm}
\begin{abstract}
A propagation torsion model for quantized vortices is proposed.The model is applied to superfluids and liquid Helium II.
\end{abstract}
ce{2cm} In a recent paper\cite{1} it was shown following earlier 
works\cite{2,3} that the magnetic field associated with the torsion induced
defects in spacetime with intrinsic spin density source could lead to line
defects carrying a quantized flux with a relation satisfying
\begin{equation}
\int QdS \cong \frac{\hbar}{ec}\ \sqrt{(9/32)(G/\alpha)}
\label{1}
\end{equation}

Here Q is the torsion vector, (connected to the spin density $\sigma$ through
$ Q = (4\pi G/(c)^{2}\sigma)\ \alpha $ is the electromagnetic fine 
structure constant, $\alpha = e^{2}/\hbar c \cong 1/137.G$ is the Newtonian 
gravitational constant. The quantity on the r.h.s. has the dimensions of
length and involves a new fundamental length different from the Planck length.
This new length involves the fundamental constants $\hbar, e, c,$ and G. In
this paper we extend the above approach to include also quantized vortices 
carrying the quantum of circulation $\hbar/m$. Earlier we had the analogy 
with type II superconductores, with the line defects induced by torsion 
carrying the flux quantum $\hbar c/2e$.

No we will have the analogy with liquid helium II, where the superfluid has
line vortices carrying the quantum of circulation $\hbar/m$.
\newpage
\pagenumbering{arabic}

In the earlie work we had the magnetic field associated with the
torsion induced line defect.

Here we would consider the torsion to be propagating and consider it
as the derivative of a scalar field thus we can write\cite{4},
\begin{equation}
Q = \nabla \phi
\label{2}
\end{equation}
Thus $\oint Q\ dS$, where dS is the surface of integration, can be written:
\begin{equation}
\oint QdS = \oint \nabla \phi \ dS = \int \nabla \ ( \nabla \phi )\ dV,
\label{3}
\end{equation}
where dV is some volume element enclosed by the surface dS.
Eq.(\ref{3}) follows from the Gauss-Ostrogradsky theorem, and can be
written :
\begin{equation}
\oint QdS \cong \int {\nabla}^{2} \phi dV
\label{4}
\end{equation}
But from the Poisson equation :
\begin{equation}
{\nabla}^{2} \phi \cong 4\pi G \rho
\label{5}
\end{equation}
(as the source of $\phi$ is the trace of the energy-momentum tensor,
the density $\rho$ in this case).

\noindent
Thus
\begin{equation}
\int {\nabla}^{2} \phi dV  \cong 4\pi G M
\label{6}
\end{equation}
where M is the total mass contained in the vortex. Since the source
of Q is the intrinsic spin, quantized in units of $\hbar$, it follows
that GM is quantized in units of $\hbar c /M (GM^{2}/\hbar c$ being 
dimensionless). Also $\oint QdS$ must have dimensions of lenght and
thus quantized as :
\begin{equation}
\oint QdS \cong n\hbar c / M
\label{7}
\end{equation}

Thus analogous to the chage flux quantum $\hbar c/ e$, we have the
mass flux $\hbar c/M$.
\newpage

Thus torsion also induces vortices carrying the circulation quantum
$\hbar c/M$. This is analogous to the situation in superfluid helium,
where quantized vortices carrying this mass flux quantum exist. The
Helmoholtz vortex theorem :
\begin{equation}
\frac{d}{dt}\  \int curl V dS = 0
\label{8}
\end{equation}
(V being the velocity potential )  implies the persistence of these
vortices in the superfluid. We have on the one hand the analogy
between torsion and the magnetic field (Q $\rightarrow B, S.Q $ the spin
torsion energy being the analogue of $\mu \ B$ the magnetic dipole 
energy\cite{3}) and the analogy between B and the vorticity field
given by $W = curl V $(i.e $B = curl A$ is analogous to $W = curl V$)
on the other hand. The torsion contact interaction\cite{2,5} can be
represented by a Heisenberg type nonlinear equation\cite{2,6}, which
can be shown to reduce to a nonlinear Schr$\ddot{o}$dinger equation
of the type known\cite{6} from the Landau-Ginzburg theory of
superfluidity. The formal analogy between the Schr$\ddot{o}$dinger
equation :
\begin{equation}
\partial \psi / \partial t = i\hbar /2m_{p}\ {\nabla}^{2} \psi
\label{9}
\end{equation}
and the Navier-Stokes viscous dissipation equation :
\begin{equation}
\partial V / \partial t = \nu \ {\nabla}^{2} V
\label{10}
\end{equation}
(where $\nu$ is the Kinematic viscosity ), implies that the quantum
viscosity of these vortices is given by
\begin{equation}
{\nu}_{p} = \hbar / 2m_{p}
\label{11}
\end{equation}
The life- time of a vortex length L is then :
\begin{equation}
t = L^{2}/ {\nu}_{p}
\label{12}
\end{equation}
\newpage

For a quantum vortex (as suggested by eq. (\ref{6}) and (\ref{7}) $L
\cong \hbar /Mc$, therefore giving
\begin{equation}
t \cong \hbar /Mc^{2}
\label{13}
\end{equation}
(from eqs. (\ref{11}) and \ref{12}).

\noindent
Eq.(\ref{13}) agrees with the uncrertainty principle estimate of the
lifetime of an unstable state of mass M, enabling the possibility of picturing
particles as quantum line vortices induced by the background torsion
geometry. It is further amusing to note that if
\begin{equation}
L \cong GM / c^{2}
\label{14}
\end{equation}
(as appropriated for describing collapsed gravitating configurations)
then the background quantum viscosity given by eq. (\ref{11}) would
imply through eqs. (\ref{12}) and (\ref{14}), a lifetime :
\begin{equation}
t \cong G^{2}M^{3}/\hbar c^{4}
\label{15}
\end{equation}
for a collapsed object of mass M. Eq. (\ref{15}) is precisely Hawking
a formula for the quantum evaporation lifetime of black hole !

The appearence of $\hbar$ in eq. (\ref{15}) suggests that the decay
is a quantum phenomenon, i.e., $\hbar \rightarrow$ o, the lifetime is
infinite. Thus quantum vortices of lenght L and mass M, if generated
in a superfluid for instance, should have a lifetime of
\begin{equation}
t \cong ML^{2}/\hbar
\label{16}
\end{equation}
This could be possibly verified in laboratory expriments. Moreover if
a particle or a beam of particles move around such a vortex line
defect eqs.(\ref{2})-(\ref{4}), suggest that it should acquire
a phase given by :
\begin{equation}
\theta = 2\pi - \oint QdS
\label{17}
\end{equation}
(torsion being associated with the nonclosure of the circuit)

In the case of strings, this so called defect angle is related to the
string tension $\mu $ as 
\begin{equation}
\theta = 2\pi (1 - 4\pi G\mu )
\label{18}
\end{equation}
This was used in ref. 7 suggest that torsion was the basis for string
tension, combining eqs.(\ref{17}) and (\ref{18}). Further analogies
will be discussed elsewhere.

We would like to thank CNPq (Brazil) for financial support as well a
Universidade do Estado do Rio de Janeiro for financial and Logistic support.

\end{document}